\documentclass[12pt]{book}

\usepackage[dvips]{graphicx,color}
\usepackage{makeidx,tsukuba}

\makeauthorindex
\makeindex

\begin{document}
\BookTitle{\itshape The 28th International Cosmic Ray Conference}
\CopyRight{\copyright 2003 by Universal Academy Press, Inc.}
\pagenumbering{arabic}
\chapter{The Lateral Distribution Function of Shower Signals in the
  Surface Detector of the Pierre Auger Observatory}

\author{Markus Roth$^1$ for the Auger Collaboration$^2$\\
  {\it (1) IEKP, Universit\"at Karlsruhe,
    POB 3640, D-76021 Karlsruhe, Germany }\\   
  {\it (2) Observatorio Pierre Auger, Malarg\"ue, 5613 Mendoza, Argentina}
}
\section*{Abstract}
The surface detector (SD) array of the southern Pierre
Auger Observatory will consist of a triangular grid of 1600 water Cherenkov
tanks with 1.5 km spacing. For zenith angles $\theta < 60^\circ$ the primary
energy can be estimated from the signal S(1000) at a distance of
about 1000~m from the shower axis, solely on basis of SD data. A suitable
lateral distribution function (LDF) S(r) is fitted to the signals recorded
by the water tanks and used to quantify S(1000). 
Therefore,  knowledge of the
LDF is a fundamental requirement for determining the energy of the primary
particle.  The Engineering Array (EA), a prototype facility consisting of 32
tanks, has taken data continuously since late 2001.  On the basis of selected
experimental
 data and Monte Carlo simulations various preliminary LDFs are
examined.  
%
\section{Introduction}
High energy cosmic rays (CRs) are detected via the extensive air showers (EAS)
they produce in the Earth atmosphere. Direction ($\theta$,$\phi$), energy (E)
and mass of the primary CR are reconstructed from the secondary particles in
the shower.  The arrival times of shower particles at various detector
locations give information on the arrival direction. The overall number of
secondaries at observation level scales roughly with primary energy, and the
form of the shower depends to some extent on the primary mass. In the Auger
experiment the longitudinal shower development is measured by the Fluorescence
Detectors (FD) while the lateral distribution at ground level is recorded by
the SD, providing two independent measurements of the shower geometry and
primary energy. The event reconstruction is hampered by the coarse sampling of
the shower particles and by the statistical fluctuations of the shower
development.  High-developing showers are expected to have a flat lateral
distribution, low-developing showers produce steeper lateral distributions.
Fortunately, at about 1 km core distance the signal is virtually independent
of primary mass and shower fluctuations, and is a good measure for the primary
energy. Thus, the energy reconstruction requires, as a first and crucial step,
to estimate S(1000) from few measured signals at various distances from an
{\em a priori} unknown core position. A second step is then to determine E
from S(1000), which relies to a large extent on shower simulations and is
therefore model dependent~[1]. We do not discuss the energy
calibration of the SD here.  Since most of the Auger events have rather few SD
stations hit, the reconstruction of the shower is not trivial. The functional
form of an LDF, S(r), and its parameters, varies with $\theta$, energy and
mass, and its determination requires a good estimate of the core position,
which in turn requires a reliable reconstruction of the shower direction,
which relies on a precise time measurement and stable trigger performance. The
estimation of S(1000) (and thus E) can be greatly improved if the shape of the
LDF is known. Here we present various approaches to determine the
LDF from experimental data and MC simulations.
\section{Probing various LDFs}
\begin{figure}[b]
\begin{center}
\vspace{-4mm}
\includegraphics[height=3.5cm]{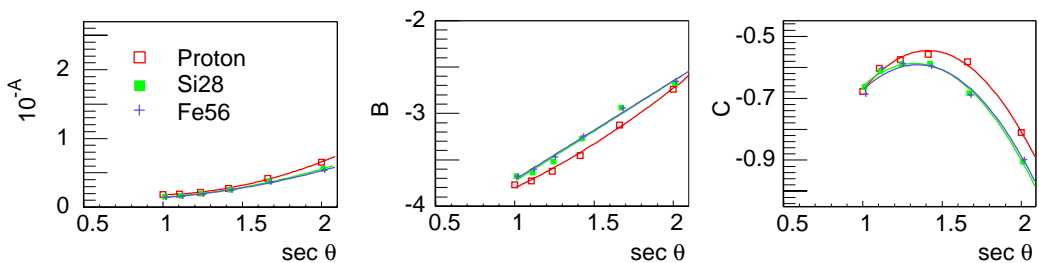}
\end{center}
\vspace{-6mm}
\caption{\footnotesize Shape parameters of LDF as function of 
  $\sec\theta$ as predicted by MC simulations~[2].}
\label{fig-pb}
\end{figure}

In contrast to S(1000) the shape of the lateral distribution does not change
much with energy. Therefore, it makes sense to decouple the normalisation
constant from the shape parameters of an LDF and to combine showers of
different energies.  Simulations of EAS with AIRES/QGSJET01 in the range E =
1-100 EeV and for $\theta = 0$-$60^\circ$ have been performed for the Auger
experiment and their output was processed through a response simulation of
detectors at core distances 200-2500 m~[2]. The LDF, in units of
vertical equivalent muons (VEM), was parametrised with an empirical function
of the form $S(r) = {\rm E}^{0.95} \cdot 10^{A+Bx+Cx^2}$ with $x =\lg(r/\rm
1000~m)$ and the parameters A, B and C were determined as function of
$\theta$ (see fig.~\ref{fig-pb}).
Independent from simulations, the LDF was also deduced from experimental data.
The EA was operating in very stable conditions during the period May to
November 2002. Therefore data from this period have been used for the
following analysis.  High-quality events have been selected, which had a
successful directional reconstruction with $\theta < 60^\circ$, signals above
3 VEM in at least 6 stations and a core position inside the EA.
High-multiplicity events are very rare: only $3\times 10^{-3}$ of the events
have 6 or more stations above threshold. Each event was examined and events or stations
with obvious problems were removed from the sample. A few well defined
events are better to determine the LDF than many events of lower quality.
The following LDFs have been investigated:
(i) a simple power law 
$ S(r) = S(1000) \cdot (r/{\rm 1000~m})^{-\nu}$ 
with a $\theta$ dependent index
$\nu = a + b \sec\theta$,
(ii) an NKG-type function:
$ S(r) = {\rm const.} \ (r/r_s)^{-\beta-0.2}\cdot (1+r/r_s)^{-\beta}
= S(1000) \cdot (r/{\rm 1000~m})^{-\beta-0.2}
          \cdot ((r+r_s)/({\rm 1000~m}+r_s))^{-\beta}$
with $\beta = a + b \sec\theta$ and $r_s = 
700$ m\footnote{Since $\beta$ and $r_s$ are strongly
correlated, we have fixed $r_s = 700$ m and left $\beta$ to vary.}, 
and (iii) the MC inspired LDF $S(r) = 10^{A+Bx+Cx^2}$.
These forms were fitted to individual events using a maximum likelihood fit of
core location and LDF at the same time. Silent (i.e. alive but no signal above
threshold) and saturated stations are properly included in the fit. The error
$\sigma(S)/S = \sqrt{0.08^2 + 0.6/S}$
of a signal S (in VEM) is taken from an analysis~[4] of data from a
closely positioned detector pair. For
\begin{figure}
\begin{center}
\includegraphics[width=0.49\textwidth]{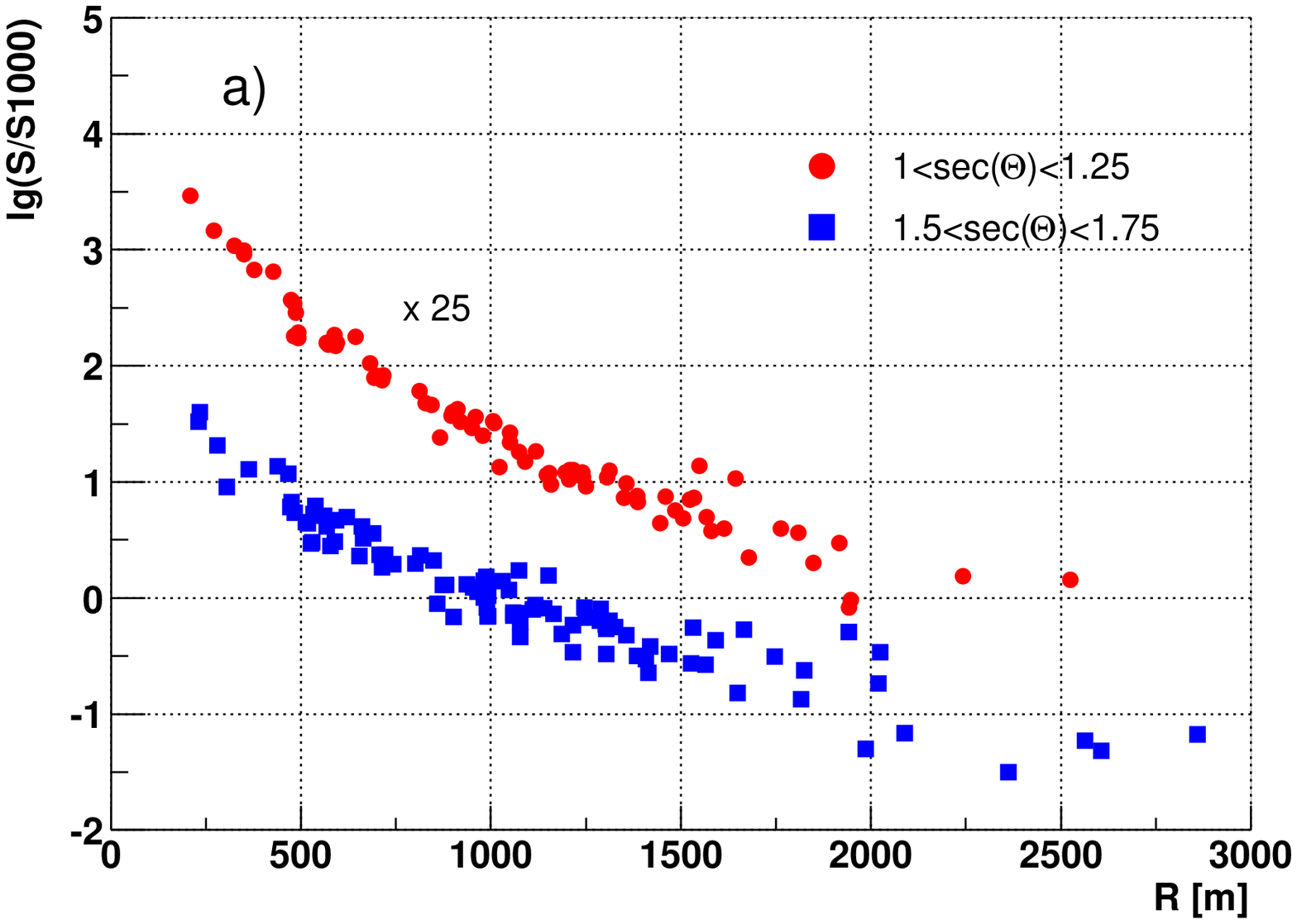}
\hfill
\includegraphics[width=0.49\textwidth]{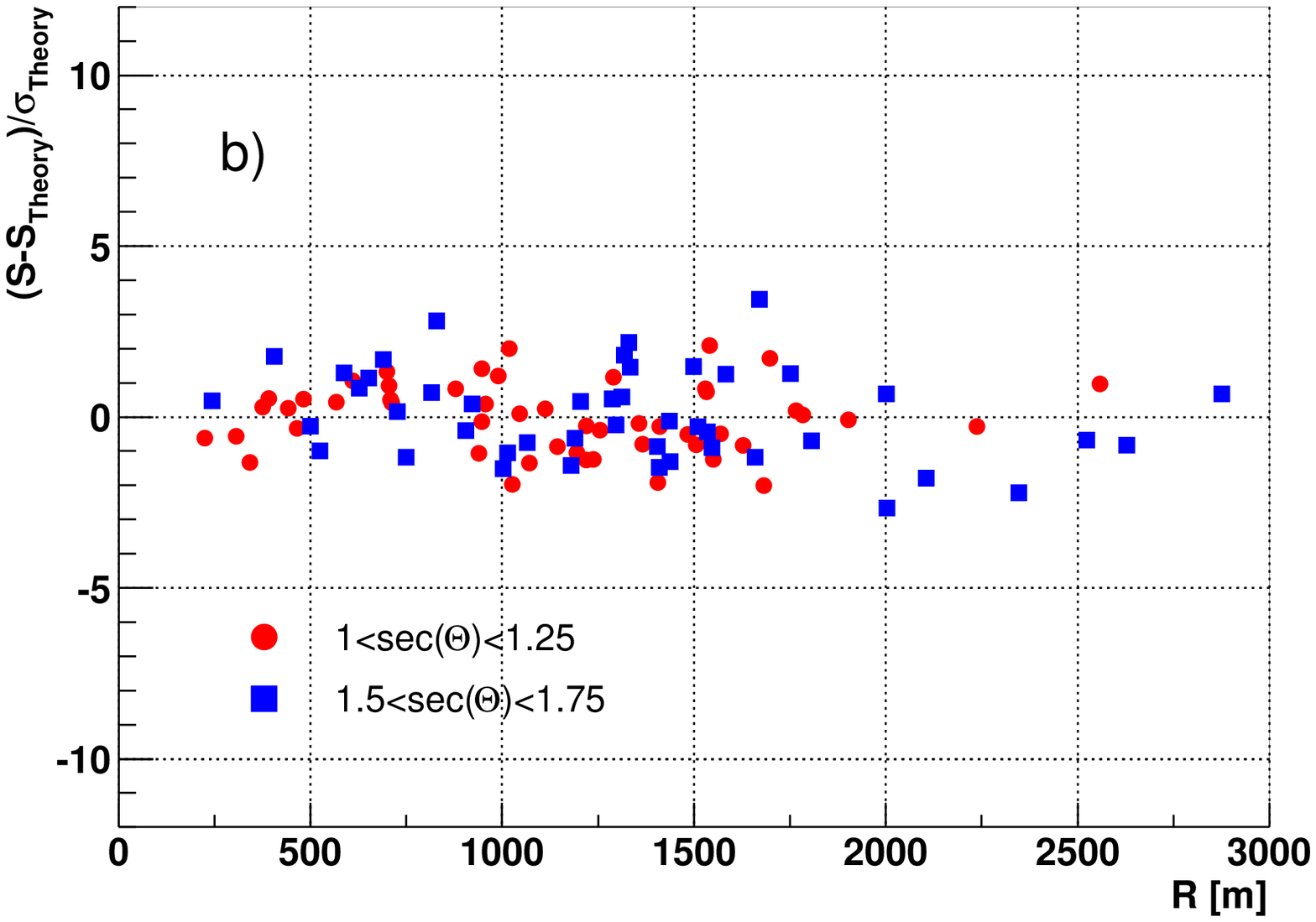}

\noindent
\includegraphics[width=0.48\textwidth]{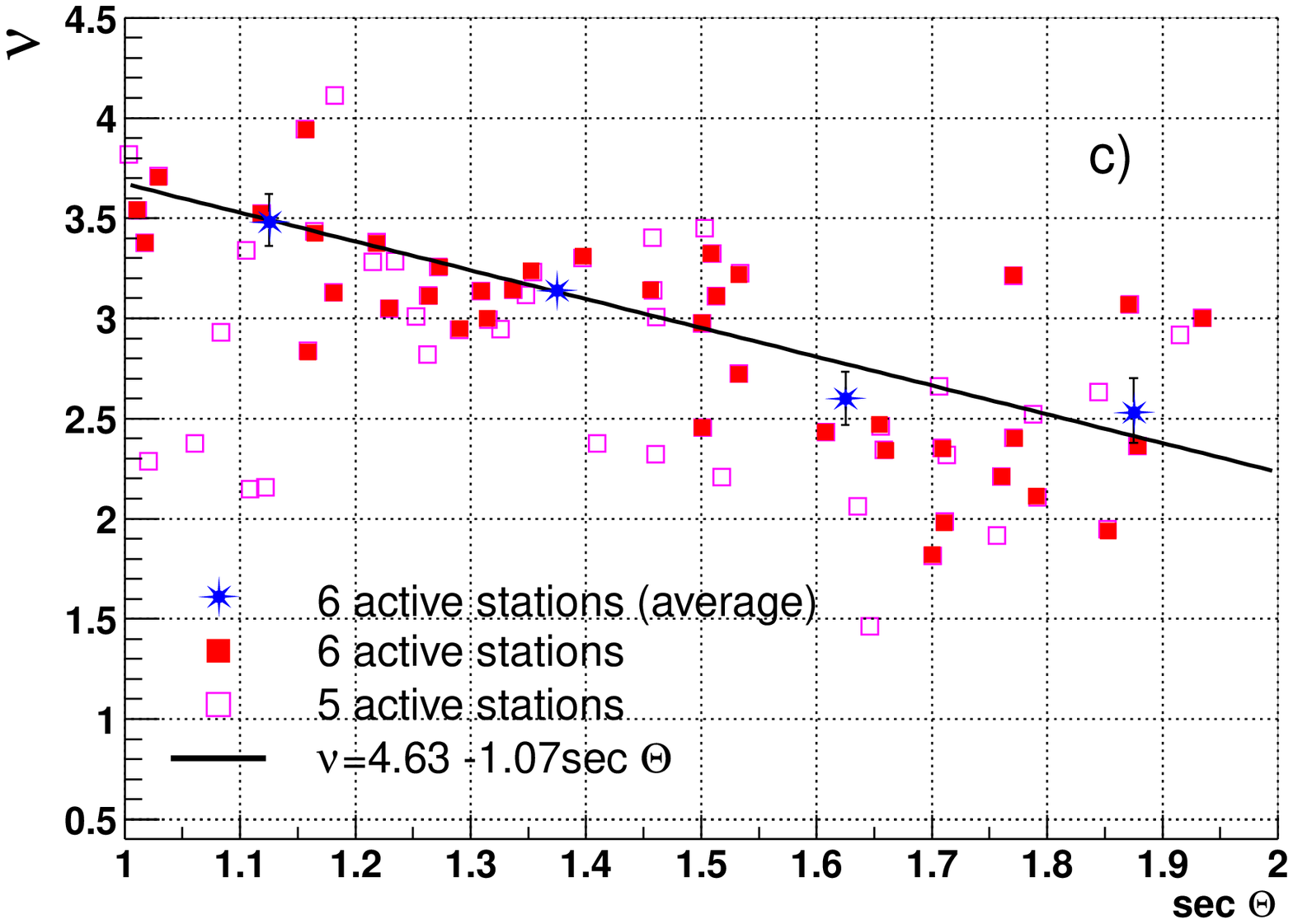}
\hfill 
\includegraphics[width=0.50\textwidth]{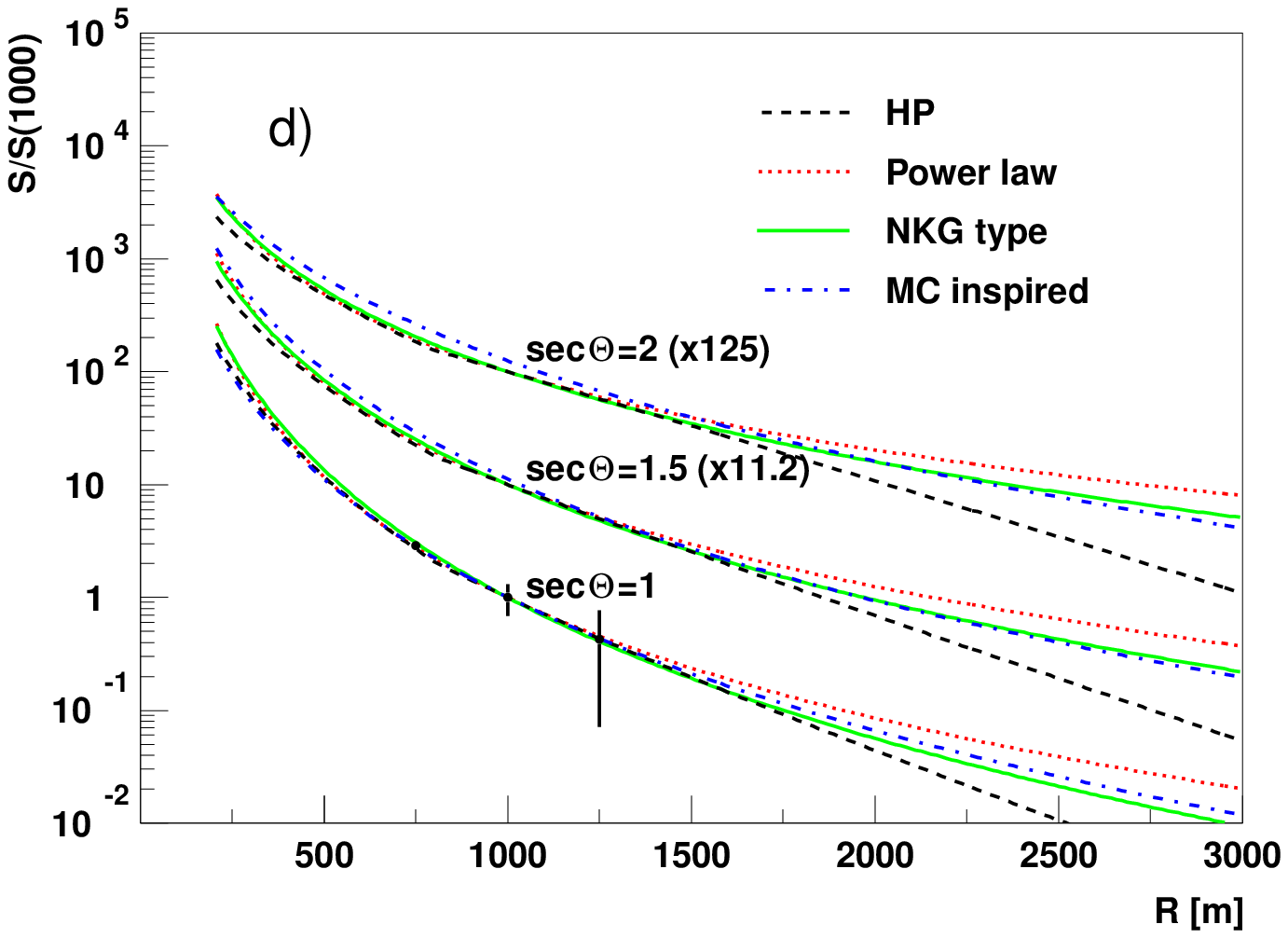}
\end{center}
\vspace{-5mm}
\caption{\footnotesize a) Measured LDF for two zenith angles (for power law
  fit).  Full symbols: stations with signal, open symbols: silent
  stations.  The data for $1 < \sec\theta < 1.25$ have been shifted upwards
  for clarity.  \quad b) Residuals $(S - S_{\rm th})/\sigma_{\rm th}$ as
  function of $r$ for two values of $\theta$ (for power law fit).  Open
  symbols are silent stations.  \quad c) Fitted value of $\nu$ as function of
  $\sec\theta$ for individual events (squares) and averaged (stars). Solid
  line: 
  fit to stars. \quad d) Different LDFs for three zenith angles. For
  $\sec\theta = 1$ a few error bars are plotted.\vspace{-5mm}}
\label{fig-ldf}
\end{figure}
the power law and the NKG-type LDFs two analyses were performed.  First, in a
two-parameter fit, the slope parameters $\nu$ and $\beta$, resp., have been
varied together with the scale factor S(1000). Then a
parameterisation of $\nu$ and $\beta$ as function of $\theta$ was determined,
which was then used in a second analysis with only fitting S(1000).
Fig. \ref{fig-ldf}a) shows the measured LDF (divided by S(1000)) for two
zenith angles when the power law assumption is used for the core finding.  To
quantify the quality of the fit, residuals $(S - S_{\rm th})/\sigma_{\rm th}$
as a function of $r$ are formed (see fig. \ref{fig-ldf}b)).  For a good LDF
the residuals should scatter for all $r$ symmetrically around 0 with a
variance of 1.  Means and standard deviations of the distribution of the
residuals are used to compare different LDFs. For the power law fit the fitted
values on $\nu$ are shown as function of $\theta$ in fig.  \ref{fig-ldf}c).
It is evident that higher-multiplicity events have a smaller scatter. A line
is fitted to the averages (stars) of the full symbols that yields $\nu =
5.1 (\pm 0.4) -1.4 (\pm 0.2) \sec\theta$. With the $\theta$  dependence of
$\nu$ 
fixed, the only fit parameter left for the LDF fit is S(1000), leading to more
stable fit results for low multiplicity events. In the same way for the NKG
function the variation of $\beta$ was found: $\beta = 3.3 (\pm 0.2) - 0.9 (\pm
0.2) \sec\theta$.  Fig.~\ref{fig-ldf}d) shows that the three chosen LDFs, with
suitably adapted parameters, agree well within the experimental errors.  For
completeness also the Haverah Park LDF~[3] is shown, that predicts
smaller densities at large core distances.
\section{Results}
The moments of the residual distributions from experimental data
including silent stations obtained with
different LDFs are listed in table \ref{tab-res} There is no major bias
apparent and all the distributions have about the same widths, which indicates
that all three LDFs describe well the data presently available.
\begin{table}
\begin{center}
\footnotesize
\hspace*{-12pt}
\begin{tabular}{|c|c|c||c|c|c|c|c|c|c|c|c|c|}
\hline
 $\sec\theta$  & 
\multicolumn{2}{|c||}{N$_{\rm evt}$ with} &
\multicolumn{4}{|c|}{power law} &
\multicolumn{4}{|c|}{NKG} &
\multicolumn{2}{|c|}{MC} \\
range &
\multicolumn{2}{|c||}{\# stations} &  
\multicolumn{2}{|c|}{$\nu$ free} &  
\multicolumn{2}{|c|}{$\nu$ fixed} &  
\multicolumn{2}{|c|}{$\beta$ free} &  
\multicolumn{2}{|c|}{$\beta$ fixed} &  
\multicolumn{2}{|c|}{ } \\  
   & $\ge 5$ & $\ge 6$& 
 m & $\sigma$& m & $\sigma$& m & $\sigma$& m & $\sigma$& m &
           $\sigma$ \\
\hline
\hline
$[1.00 , 1.25]$&21&11& -0.05 & 0.50 & -0.11 & 0.65 & -0.03 & 0.49 & -0.05 & 0.57 & -0.14 & 0.66 \\
$[1.25 , 1.50]$&18&9& -0.07 & 0.61 & -0.07 & 0.62 & -0.02 & 0.53 & -0.05 & 0.57 & 0.06 & 0.93 \\
$[1.50 , 1.75]$&18&12& -0.05 & 0.65 & -0.04 & 0.73 & -0.03 & 0.63 & -0.02 & 0.71 & 0.02 & 0.80 \\
$[1.75 , 2.00]$&12&8& -0.08 & 0.83 & -0.15 & 1.15 & -0.11 & 1.13 & -0.10 & 1.17 & -0.15 & 1.29 \\
\hline
\end{tabular}
\end{center}
\vspace{-3mm}
\caption{\footnotesize Moments (mean, $\sigma$) of residual distribution of exp. data 
with various LDFs. Only events with $\ge 6$ stations were used in the present analysis.
\vspace{-5mm}}
\label{tab-res}
\end{table}
Previous experiments have shown that a pure power law cannot describe the
shower signals at large core distances~[3].  The results presented here
are still preliminary as the statistics of events from the Auger EA,
especially for energies $> 10^{19}$ eV, are small and since preliminary
algorithms for directional and core reconstruction are used. In future each
improvement on the statistics, angular and core resolution, and specifically
hybrid events with their superior geometric reconstruction, will also improve
the knowledge on the LDF and permit finer details to be analysed.
%
\vspace{\baselineskip}
\re
1.\ Billoir P., Auger Collaboration internal note, GAP 2002-075
\re
2.\ Billoir P., Da Silva P.,  Auger Collaboration internal note, GAP 2002-073
\re
3.\ Coy R.N. et al., Astrop. Phys. 6 (1997) 263
\re
4.\ Yamamoto T. for the Auger Collaboration, these proceedings
%
\endofpaper
\end{document}